\documentstyle[psfig]{mn}

\def\fd{\hbox{$.\!\!^{\rm d}$}}

\title[RX\, J1039.7-0507: A New Intermediate Polar and Probable Recent Nova]
{RX\, J1039.7-0507: A New Intermediate Polar and Probable Recent Nova, possessing a large Reflection Effect}
\author[Patrick A. Woudt and Brian Warner]
       {Patrick A. Woudt\thanks{E-mail: pwoudt@circinus.ast.uct.ac.za} 
        and Brian Warner\thanks{E-mail: warner@physci.uct.ac.za}\\
        Department of Astronomy, University of Cape Town, Private Bag,
        Rondebosch 7700, South Africa}
\date{}

\begin{document}

\maketitle

\begin{abstract}
The ROSAT source RX\,J1039.7-0507, known from its optical spectrum to be a Cataclysmic
Variable star, is shown from high speed photometry to have an orbital period of 1.574 h. The
system has a nearly sinusoidal photometric modulation with a range of 1.1 mag which we interpret
as the reflection effect caused by a very hot white dwarf primary. This suggests that a
nova explosion occured on the primary in the relatively recent past.
In addition, RX\,J1039.7-0507 has periodic signals at 1932.5 s and 721.9 s which, with the aid of 
other periods present at low amplitude, we interpret as an orbital sideband and the first harmonic
of the primary's spin period. RX\,J1039.7-0507 is therefore an intermediate polar with a spin
period of 1444 s (24.07 min).
\end{abstract}


\section{Introduction}

RX\, J1039.7-0507 (hereafter RXJ1039) is a weak ROSAT X-Ray source in the constellation of 
Sextans that was identified as a V = 18.5 cataclysmic variable (CV) star by Appenzeller 
et al.~(1998) on the basis of positional agreement and an emission line spectrum. Its 
time resolved light curve has remained unknown until now. Hoard et al (2002) obtained 
$J$, $H$, $K$ photometry which revealed unusual colours, but in view of the large amplitude 
of variation that we observe, interpreted as a strong reflection effect, the use of  
colours obtained at an unknown orbital phase is not straightforward. Our high speed 
photometry has shown RXJ1039 to be a CV of unusual interest, worthy of further 
observation in other wavelength regions than the optical band photometry that we present here.

\section{Observations}

Our high speed photometry was carried out during April 2002 at the Sutherland station 
of the South African Astronomical Observatory. The University of Cape Town's CCD photometer 
(O'Donoghue 1995) was used on the 74-in and 40-in reflectors. No filter was employed, giving 
a response probably peaking in the yellow for CVs with strongly increasing flux towards 
shorter wavelengths. Reductions were made in the standard way, making use of other stars 
within the CCD frame. Some of the observations were made in non-photometric conditions, 
but this did not noticeably reduce the final quality of the light curves.

\begin{figure*}
\centerline{\hbox{\psfig{figure=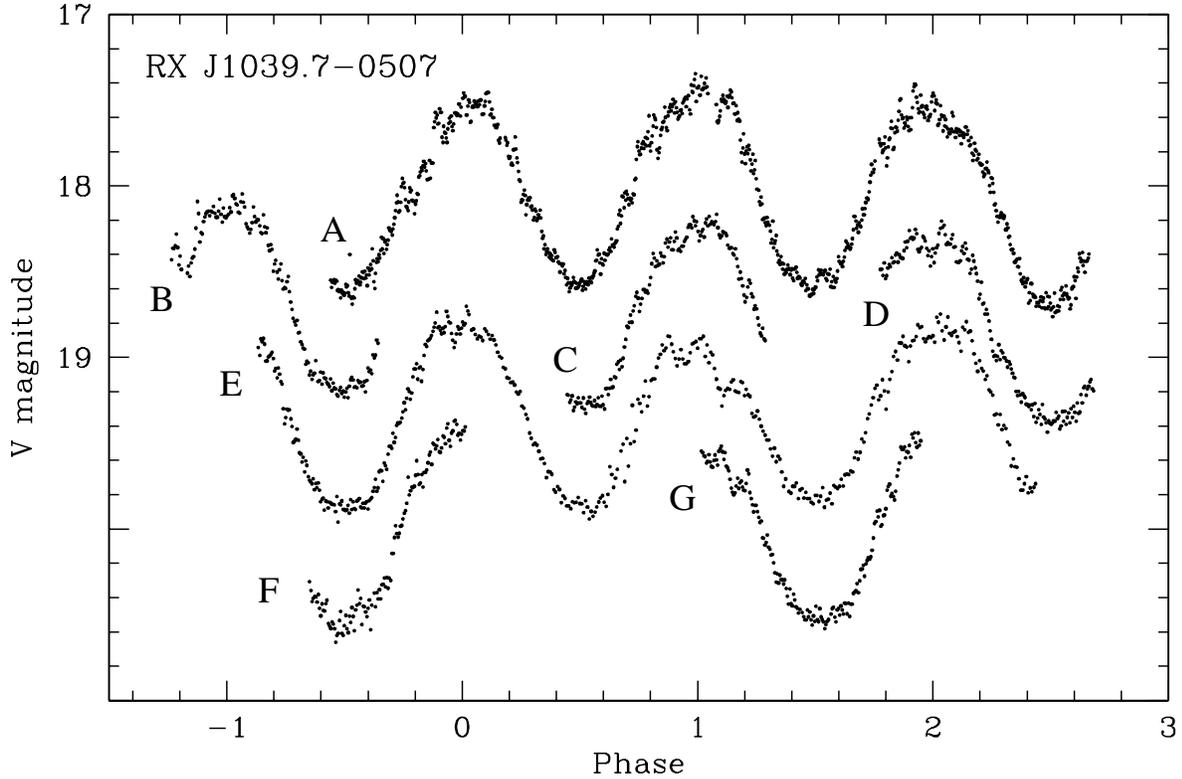,width=16cm}}}
  \caption{The light curves of RX\,J1039-0507 obtained in 2002 April, phased 
according to Eqn.~\ref{eph}. The upper light curve shows the correct brightness. Other light
curves have been shifted vertically for display purposes.}
 \label{fig1}
\end{figure*}

The observing runs and their details are listed in Table~\ref{tab1}. Initially, having 
found a large amplitude almost sinusoidal light curve, we obtained further light curves 
of short duration merely for the purpose of increasing the accuracy of the orbital period, but when 
the presence of other periods became clear in the Fourier transform (FT) we gave more 
intensive attention.
     
Figure~\ref{fig1} displays all our light curves, phased according to the ephemeris 
derived from the FT of the combined data. This ephemeris, giving the times of maximum light, is

\begin{equation}
{\rm HJD_{max}} = 2452371.331954 + 0\fd065597 (\pm 1) \, {\rm E}.
\label{eph}
\end{equation}

where the error in period (0.1 s) is obtained from a non-linear least squares fit 
(and which from experience is known greatly to underestimate the true uncertainty).

\begin{table*}
 \centering
  \caption{Observing log.}
  \begin{tabular}{@{}llllccc@{}}
 Run No.  & Date of obs.          & HJD of first obs. & Length    & $t_{in}$ & Tel. & $<$V$>$ \\
          & (start of night)      &  (+2452000.0)     & (h)       &     (s)   &      & (mag) \\[10pt]
A = S6351    & 2002 April 6  &  371.29532  &   5.07      &      20   &  74-in & 18.0\\
B = S6354    & 2002 April 7  &  372.23487  &   1.38      &      30   &  74-in & 18.1\\
C = S6365    & 2002 April 10 &  375.23141  &   1.33      &      30   &  40-in & 18.3\\
D = S6375    & 2002 April 12 &  377.22106  &   1.42      &      30   &  40-in & 18.4\\
E = S6382    & 2002 April 13 &  378.22843  &   5.20      &      30   &  40-in & 18.4\\
F            & 2002 April 18 &  383.22807  &   1.04      &      30   &  40-in & 18.4:\\
G            & 2002 April 21 &  386.28913  &   1.47      &      30   &  40-in & 18.4:\\
\end{tabular}
{\footnotesize 
\newline 
Notes: $t_{in}$ is the integration time, `:' denotes an uncertain value.\hfill}
\label{tab1}
\end{table*}

The light curve shows an almost sinusoidal intensity modulation with a period 
of 5667.6 s (1.574 h) and a peak-to-peak range of a factor of 2.73, i.e. an 
amplitude (half the range) of 0.545 mag, about a mean magnitude of 18.0. A slight 
departure from sinusoidality is seen by the existence in the FT of a first harmonic 
(which gives an independent estimate of the orbital period of 5664.1($\pm$0.7) s) with an 
amplitude of 0.043 mag. The presence of some low level flickering is consonant with 
the classification of RXJ1039 as a CV, but much of the apparent rapid variability 
is caused by the coherent periodic variations that we discuss below. At a period 
of 1.57 h RXJ1039 is clearly below the `orbital period gap' of CVs.
   
The very large amplitude of the principal photometric modulation is the signature 
of a reflection effect. There are no detectable eclipses, which implies an inclination 
less than about 65$^\circ$, but the inclination cannot be much less than this for, with 
its peak-to-peak range of 1.1 mag, RXJ1039 has among the largest amplitude modulations
observed for CVs and detached white dwarf/M dwarf binaries (see, e.g., Chen et al.~1995). 
The implication is that the primary in RXJ1039 is very hot, and this in turn suggests 
that it underwent a nova eruption within the past decade or so. The known novae with 
periods below the period gap (RW UMi, GQ Mus, CP Pup, V1974 Cyg, RS Car) include 
representatives from fast and slow classes of light curves, with eruption ranges 
around 10 - 13 mag. With a quiescent magnitude of 18.0, RXJ1039 would probably therefore 
not have reached naked eye brightness and would be most likely to have been found 
-- or missed -- in photographic surveys (which in any case have been reduced 
in completeness in the past two decades).  At $\ell = 253^\circ$, $b = 45^\circ$, 
RXJ1039 is far from the direction of the Galactic Centre, towards which most photographic 
surveys for novae have been concentrated. RXJ1039 is therefore a prime candidate for first
membership of a class of relatively recent `overlooked' novae. The ROSAT catalogue of X-Ray 
sources may contain other such waifs; their eventual identification will give an 
indication of the completeness of optical searches in the past decades.

We have searched the available data archives to see if any observations in the vicinity
of RXJ1039 were made in the past 10 years. No observations were found, although we note the
proximity ($\sim$75 arcsec) of a small galaxy (2MASXi J1039517-050649). 

\section{RXJ1039 as an Intermediate Polar}

\begin{figure}
\centerline{\hbox{\psfig{figure=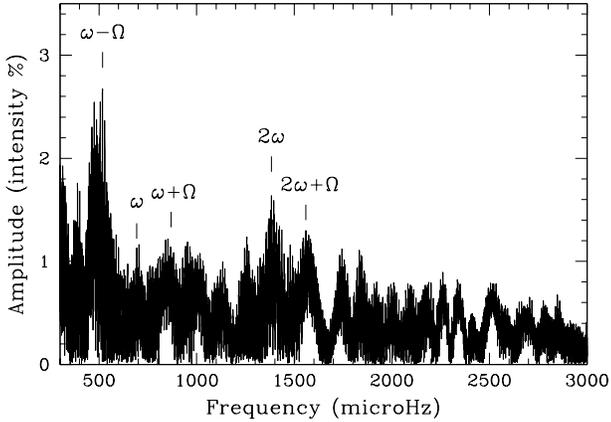,width=8.2cm}}}
  \caption{The Fourier transform of the entire set of data of RX\,J1039.7-0507. }
 \label{ftrxj1039}
\end{figure}

\begin{figure}
\centerline{\hbox{\psfig{figure=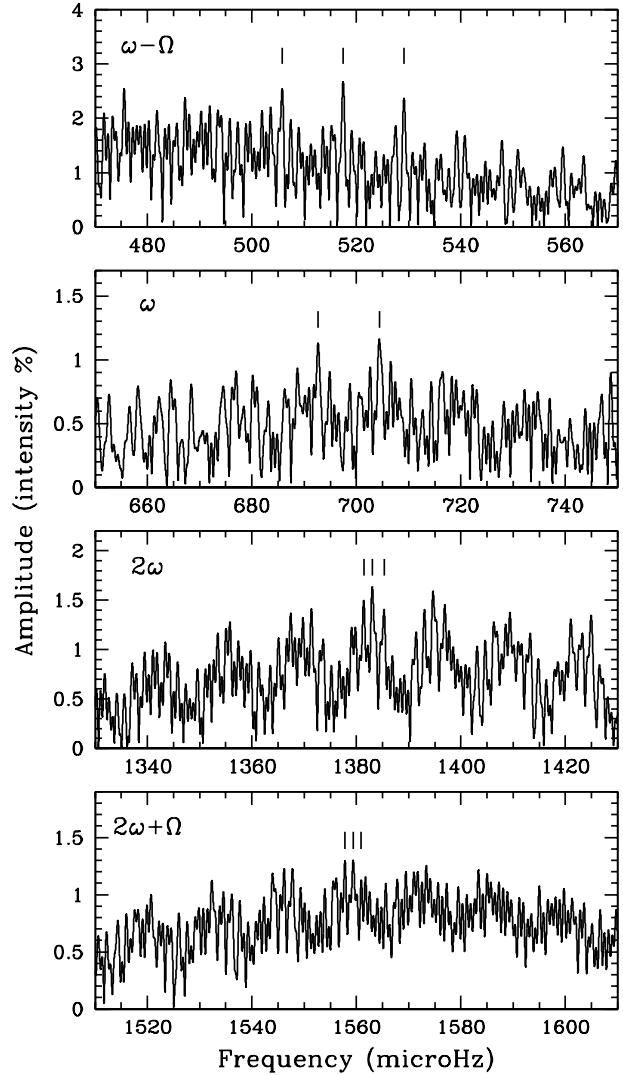,width=8.2cm}}}
  \caption{A detailed view of the Fourier transform of the entire set of data of RX\,J1039.7-0507
around the four identified frequencies associated with $\omega$, $\Omega$ and combinations thereof. 
The ordinate is fractional intensity, in percent.}
 \label{ftburxj1039}
\end{figure}

The FT (of intensities) of our entire data set (with the mean of each run subtracted), 
prewhitened at the orbital period and its first 
harmonic, is shown in Figure~\ref{ftrxj1039}. We have not displayed the lowest frequencies, where there
is a broad low amplitude feature centred on the orbital frequency which we attribute to sidebands that occur
because of small non-periodic modulation of the signal. Various combinations of runs show two 
persistent prominent peaks in the vicinity of 1933 s and 723 s. In the FT of the complete data set 
these appear at 1932.5 s with an amplitude of 2.6\%, and at 721.87 s with 
an amplitude of 1.4\%. There are also aliases to these periods -- see Table~\ref{tab2} and Figure~\ref{ftburxj1039}. 
Expressed as frequencies, the two prominent modulations are at 
$f_1$ = 517.5 $\mu$Hz and $f_2$ = 1383.0 $\mu$Hz respectively, and we notice 
that $f_2$ -- 2 $f_1$ = 348.0 $\mu$Hz $\approx 2 \Omega$, where $\Omega$ is the orbital frequency (176.44 $\mu$Hz). 
From comparison with the optical modulations seen in intermediate polars (IPs - e.g. Patterson (1994)) 
this suggests the identification of $f_1$ with the frequency $\omega - \Omega$ and $f_2$ with 2$\omega$, 
where $\omega$ is the spin period of the white dwarf primary. This receives support from the existence 
in the first two of our runs of a signal near $\omega$ which shows in the total FT as a modulation 
at 1443.7 s (i.e. 692.7 $\mu$Hz). We also find a modulation in the FT at 640.3 s 
(1562 $\mu$Hz) that we identify as a 2$\omega$ + $\Omega$ modulation. 
There is a peak of power in the FT near a $\omega$ + $\Omega$ modulation (see Figure~\ref{ftrxj1039}) but we cannot
find a convincing amplitude spike that coincides with it or its alias; there is 
no evidence in the FT for a 2$\omega$ -- $\Omega$ modulation. 
Only the set of frequencies that we have discussed here produces a sensible model -- choice 
of any of the aliases of these selected modulations does not lead
to any numerical relationships that match what is recognised as typical IP behaviour.
    
\begin{table}
 \centering
  \caption{The suite of frequencies and their aliases. Fractional intensities are given in brackets. The
selected frequencies are listed in bold.}
  \begin{tabular}{@{}cccc@{}}
ID                      & Frequency    & Frequency  & Frequency \\
                        & $\mu$Hz      & $\mu$Hz    & $\mu$Hz  \\[10pt]
$\omega - \Omega$       & 505.8 (2.54\%)  & {\bf 517.5 (2.63\%)}   &  529.1 (2.32\%)        \\
$\omega$                & {\bf 692.7 (1.14\%)}  & 704.4 (1.17\%)   &                        \\
2 $\omega$              & 1381.4 (1.49\%)       & {\bf 1383.0 (1.64\%)}  & 1385.3 (1.41\%)  \\
2 $\omega + \Omega$     & 1557.8 (1.30\%)       & {\bf 1559.4 (1.30\%)}  & 1560.9 (1.13\%)  \\
\end{tabular}
\label{tab2}
\end{table}

We have not attached uncertainties to the frequency measurements; many of the modulations do not rise
much above the noise (see Figure~\ref{ftburxj1039}) and will be influenced in an unknown and non-random
way by the noise. For our strongest ($\omega - \Omega$) modulation we find from an FT of runs A, B and C
517.4 $\mu$Hz (2.53\%) and from runs C, D and E 516.7 $\mu$Hz (3.23\%). In the weaker signals errors of
frequency of 1 $\mu$Hz or greater can therefore be expected.

This suite of modulations gives strong evidence that RXJ1039 is an IP with a spin period of 1444 s, 
accreting on to two magnetic poles, where the strongest optical signal is the orbital 
sideband at $\omega$ -- $\Omega$ and there is variable amplitude modulation at the 
spin period $\omega$ and its first harmonic.

\section{Discussion}

The large amplitude of orbital modulation caused by a reflection effect is characteristic of a recent nova in which 
the accretion disc does not dominate the luminosity of the system (Prialnik 1986; Kovetz, 
Prialnik \& Shara 1988). This requires either a short orbital period, in which the 
disc has relatively small dimensions, or a very high inclination (so that the disc is 
highly foreshortened but the irradiated secondary is seen to best advantage), or a polar, 
in which there is no disc at all. Even at short $P_{orb}$ there will only be high 
amplitude modulation for normal postnova mass transfer rates ($5-15 \times 10^{-9}$ M$_{\odot}$ y$^{-1}$)
(Prialnik 1986) if the orbital inclination is large. 

Some examples are the 
nova/polar V1500 Cyg, with $P_{orb}$ = 3.35 h, which had a modulation range 
of 1.2 mag 12 years after eruption (Kaluzny \& Chlebowski 1988) reducing to 0.7 mag
in 1995 (Somers \& Naylor 1999) and V351 Pup (Nova Puppis 1991). The latter
has an orbital modulation range of 1.1 mag which, at $P_{orb}$ = 2.84 h, 
implies a magnetic system (Woudt \& Warner 2001). On the other hand, in DN Gem (Nova Gem 1912), 
with $P_{orb}$ = 3.07 h, the primary has cooled sufficiently and the luminosity of the disc 
is so high that the irradiated secondary produces a modulation range of only 0.06 mag (Retter, Leibowitz
\& Naylor 1999); but in V1974 Cyg (a nova in 1992), which has $P_{orb}$ = 1.95 h and might 
have been expected to have a larger modulation range, the range is only 0.05 mag in $V$ 
(DeYoung \& Schmidt 1994) because the inclination is quite low (38$^\circ$: Chochol et al 1997). 
In addition, we have recently found DD Cir (Nova Circini 1999) to have an orbital period
of 2.34 h and a reflection effect with a range of 0.20 mag. An anomalous
example is the recurrent nova V394 CrA, which has $P_{orb}$ = 18.2 h 
(Schaefer 1990) but manages to exhibit a 1.0 mag reflection effect that can be explained 
by irradiation caused by the very high rate of mass transfer ($1.5 \times 10^{-7}$ M$_{\odot}$ y$^{-1}$)
and extreme mass (1.37 M$_{\odot}$) of the primary which generates 
an exceptionally high accretion luminosity (Hachisu \& Kato 2000).

RXJ1039 is a good candidate for an extended pointed observation in the X-ray region,
and for spectra to be taken in the satellite ultraviolet. It evidently has a very hot
white dwarf that is causing the reflection effect. This and its apparent magnetic nature
should give a large UV and soft X-ray flux, which will be modulated at the rotational
period.

\section*{Acknowledgments}

The authors kindly acknowledge Dr.~E.~Romero-Colmenero for the additional observations
in 2002 April. Furthermore, we thank Dr.~G.~Israel for searching the ROSAT database on our request.
BW is funded by the University of Cape Town, PAW is funded by the NRF and partly through strategic
funds made available to BW by the University of Cape Town.

\end{document}